\documentstyle[12pt,a4,epsfig,wrapfig]{article}

\renewcommand\thesection{\arabic{section}.}
\renewcommand\thesubsection{\thesection\arabic{subsection}.}
\renewcommand\thesubsubsection{\thesubsection\arabic{subsubsection}.}
\renewcommand\section[1]{\vspace{\topsep}\vspace{\partopsep}
\refstepcounter{section}
{\par  \noindent\normalsize\bf \thesection
\hspace{1em}#1\vspace{\topsep}\par\noindent}}

\newenvironment{refs}
{\vspace{\topsep}\vspace{\partopsep}
{\par \noindent\normalsize\bf  References
\vspace{-\topsep}\par\noindent}
\setlength{\parindent}{-5mm}
\begin{list}{}{\topsep 0pt \partopsep 0pt \itemsep 0pt \leftmargin 5mm
\parsep 0pt \itemindent -5mm}}
{\end{list}}

\renewcommand\subsection[1]{
\refstepcounter{subsection}
{\par \protect\vspace{\topsep}\vspace{\partopsep}
 \noindent\normalsize\bf \it \thesubsection
\hspace{1em}#1\par \noindent}}

\renewcommand\subsubsection[1]{
\refstepcounter{subsubsection}
{\par \protect \vspace{\topsep}\vspace{\partopsep}
\noindent\normalsize \it \thesubsubsection
\hspace{1em}#1\par \noindent}}

\newfont{\sansb}{cmssbx10}
\newfont{\sans}{cmss10}

\begin{document}
\begin{center}
{\large \bf On an extragalactic origin of the 
dominant part of the hadronic cosmic rays\vspace{18pt}\\}
{ R.Plaga$^a$\footnote{email address:
        plaga@hegra1.mppmu.mpg.de}\vspace{12pt}\\}
{\sl
{\it $^a$Max-Planck-Institut f\"ur Physik, F\"ohringer Ring 6,
        D-80805 M\"unchen, Germany  \vspace{12pt}\\}
}
\end{center}
\begin{abstract}
The possibility that the major part of all extrasolar
hadronic cosmic rays with energies above 10 MeV/n
is of extragalactic origin is discussed.
Recent observational 
results on the galactocentric cosmic-ray density
gradient and very high $\gamma$-ray emission 
do not support expectations
from the simplest models with a Galactic origin of
cosmic rays.
The hypothesis that ``flux trapping'' of extragalactical
cosmic-rays occurs in the Galactic confinement volume
is advanced. Taking this phenomenon into account, all
the usual objections against an extragalactic origin of
{\it hadronic} cosmic rays loose their strength. 
The local energy density
of hadronic cosmic rays and other observational facts
can be understood in a very natural way assuming
an extragalactical origin.
A promising scenario seems to be a
Galactic origin of {\it electrons} and an extragalactic
origin of {\it hadrons}.
\end{abstract}
\setlength{\parindent}{1cm}
\section{Introduction}
This report is a short summary and update 
of a more detailed article
(Plaga, 1997). In addition, in the third part counterarguments
against an extragalactic origin of cosmic rays are 
explicitely answered.
The possibility of an extragalactic origin
of the dominant part of all extrasolar {\it hadronic} cosmic rays
has been considered since the early days
of cosmic-ray research (Baade and Zwicky, 1934). 
In the modern era it was discussed by Burbidge, Hoyle and 
Brecher (see e.g. Brecher and Burbidge, 1972; Burbidge, 1974).
It has long been clear that the electron component
of cosmic-rays is of Galactic origin, and recently there
has been new impressive evidence for a supernova
origin of this component from TeV astrophysics (Tanimori et al,1997;
T.Kifune, these proceedings).
\\
Since about 30 years, using arguments which I will
discuss below, the standard view about the origin of
extrasolar cosmic rays has been that they must be produced
in our Galaxy. In this case the only objects which seem
to be capable to sustain the observed cosmic-ray energy
density in the Galaxy against losses to intergalactic space are
supernova remnants (Berezinskii et al.,1990).
This idea is eminently plausible because shock-wave
acceleration is expected to operate in these objects
(Berezhko and V\"olk, 1997; H.V\"olk, these proceedings).
There are two observational facts, however, which seem
not to be in complete agreement with the most
simple quantitative models of a Galactic origin of the main part
{\it hadronic} cosmic rays:
\\
1. A classical problem in cosmic-ray physics and
$\gamma$-ray astrophysics is to infer 
cosmic-ray density as a function
of the distance from the Galactic center.
\begin{figure}[hbt]
\centering
\epsfig{file=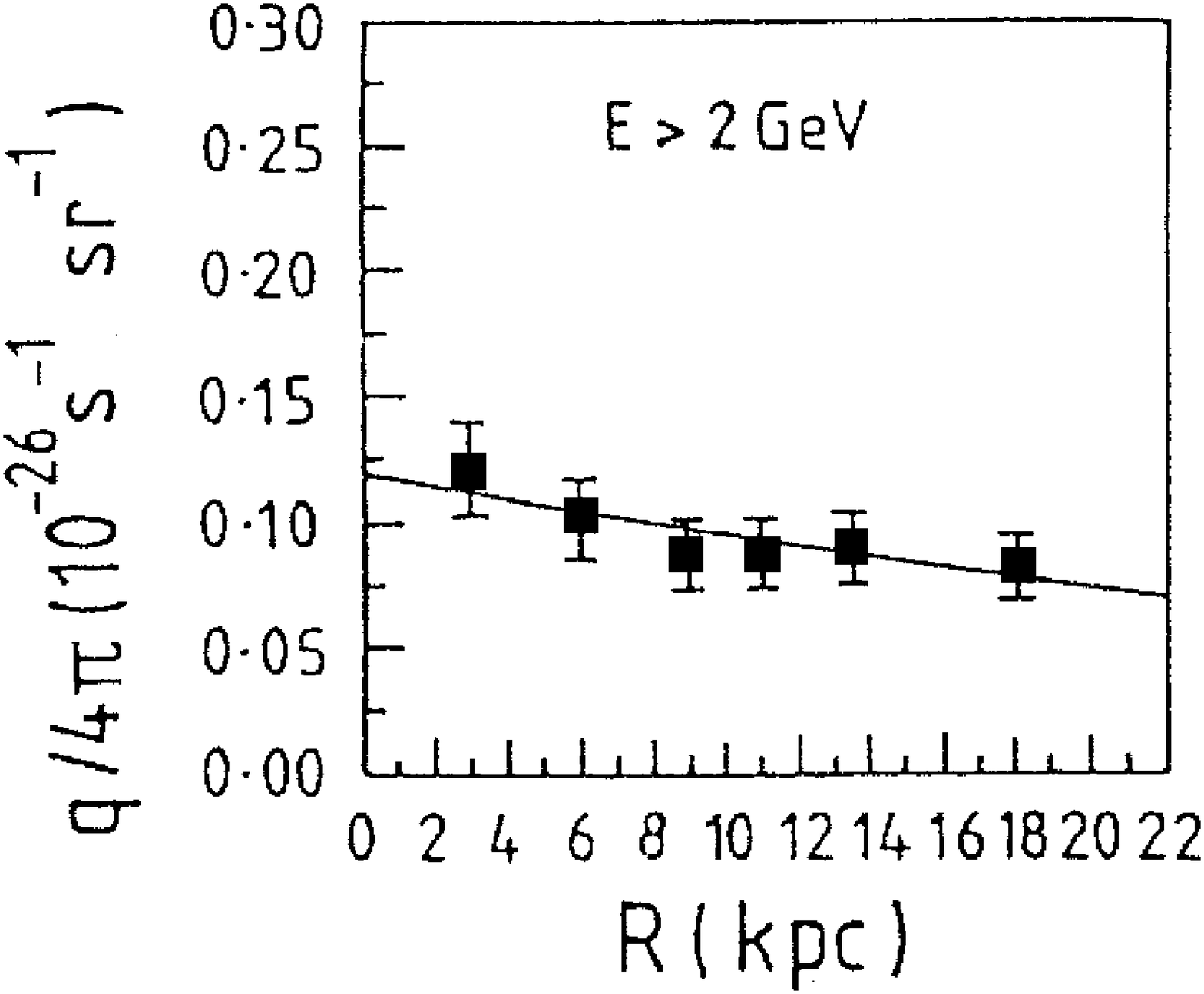,width=9cm,height=5cm,clip=,angle=0}
\caption{ \label{rad_grad} The figure 
(directly taken from (Erlykin et al.,1997)) 
shows the $\gamma$-ray emissivity
(proportional to cosmic-ray intensity) versus Galactocentric
radius inferred from EGRET data. }
\end{figure} 
This is done by measuring the $\gamma$-ray intensity
due to cosmic-ray - interstellar medium interactions.
Taking into account the known density of the interstellar
medium, the cosmic-ray density can be inferred.
Fig.1 shows the resulting density from $\gamma$-ray
data with $\gamma$-ray energies above 2 GeV from the publication
of Erlykin et al. (1996) based on EGRET data.
This result (in basic agreement
with other independent analyses) seems to indicate
that beyond about 8 kpc there is no clear 
gradient, i.e. the cosmic-ray density seems to
remain constant, or at least does not
fall by more than about 30 $\%$ between 8 and 18 kpc.
Because supernova remnants and all other
proposed Galactic accelerators are strongly concentrated towards
the Galactic center, one expects in simple diffusion models
of Galactic propagation a fall in comic-ray intensity 
by more than a factor of 10
in cosmic-ray density between 8 and 18 kpc (Case 
and Bhattacharya, 1996). 
While it is certainly possible to devise more
complicated models with an extended halo of
cosmic-rays and transport with winds rather than
diffusion (Erlykin et al., 1997) which explain this
strong disagreement, it should be noted
that {\it a constant density of cosmic-rays at
different distances from the Galactic center (as
observed), has always been 
considered as the signature for an extragalactic
origin of cosmic rays.}
The observed small gradient at distances smaller
than 8 kpc can be  naturally interpreted as due to electrons, which are
of Galactic origin.
\\
2. Another recent clue from TeV astrophysics
have been strict upper limits on $\gamma$-ray
emission from certain supernova remnants 
(He\ss \, 1997;Lessard et al.,1997).
For two special cases (SNR G78.2+2.1 and IC443) these
limits lie at the lower end of the the uncertainty range 
for $\gamma$-radiation of hadronic origin, made
under the assumption of a Galactic 
supernova origin of cosmic rays.
Though clearly these results do not signal a crisis
for the Galactic scenario for cosmic-ray origin, they
underline the fact that no {\it direct} evidence 
of any sort has been found for it up to now.
\\
The question I ask is {\it not} ``do Galactic supernova remnants
accelerate also ions in addition to electrons?''.
It seems to me very likely that they do that. The question
is rather: ``Do they accelerate them in the {\it required amount} 
(i.e. about a factor of 100 times more efficient than the
electrons)
to explain the local cosmic-ray energy density?''

\section{Magnetic flux trapping}
\vspace{-0.9cm}
\subsection{The physical idea}
The new idea of this contribution is ``magnetic-flux trapping'',
a mechanism that might lead to {\it enhancement} 
of the intergalactic
cosmic-ray density in the Galactic confinement volume
by a factor $e$.
If cosmic rays escape from an
extragalactic production  site with a high magnetic field 
$B_{\mathrm{er}}$
into intergalactic space with a low magnetic field 
${B_{\mathrm{IGM}}}$
{\it conserving
the adiabatic invariant},
they have small pitch angles below
a maximal angle
$\theta_{\mathrm{max}}$ = arcsin(
$\sqrt{B_{\mathrm{IGM}}/B_{\mathrm{er}}}$).
Under these conditions the particles can freely
\begin{figure}[hbt]
\centering
\epsfig{file=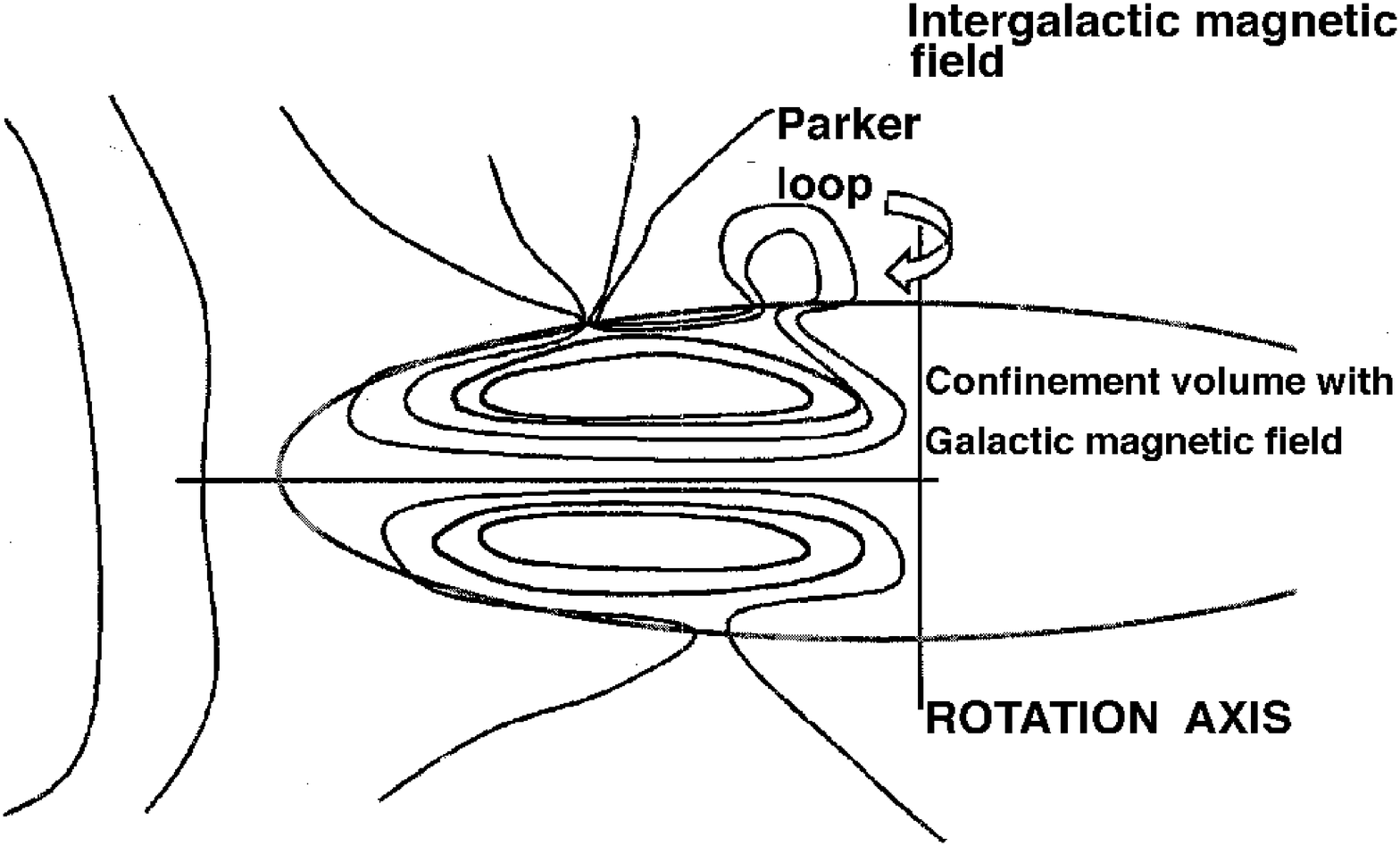,width=9cm,height=6cm,clip=,angle=0}
\caption{ \label{mag_struc}Schematic sketch of the hypothetical general
topology of the Galactic and intergalactic magnetic field
assumed for the present scenario. The field is
generally closed,
but some Parker loops connect to the intergalactic magnetic
field. Cosmic rays enter through the connected loops and
escape through all loops.}
\end{figure} 
enter the Galactic confinement 
volume via the open field lines, if the 
Galactic field has a general topology as depicted in figure
\ref{mag_struc}.
\\
It is generally agreed that the adiabatic invariant
is not conserved inside the Galaxy due to various
scattering mechanisms. All Galactic field lines
are expected to be filled with particles of all
pitch angles. 
If the only cosmic-ray loss path is via
open field lines connected to intergalactic field
lines, the cosmic ray density inside the Galaxy
is a factor
\\
(1) $b_{\mathrm{conc,equ}}=2B_{\mathrm{Gal}}/B_{\mathrm{IGM}}$
\\ 
enhanced over the intergalactic value.
Here $B_{Gal}$ is the Galactic field strength.
{\it This could lead to concentration factors on the order
of 10$^5$ with plausible values for the intergalactic and Galactic
magnetic field.} This enhancement mechanism was first
dicussed in a related context by Sciama (1962).
The main speculative parts of the hypothesis are that:
\\
1. the incoming intergalactic cosmic rays
have very small pitch angles so that they can freely
enter the Galactic confinement volume.
This will be the case if their place of origin lies in
regions with magnetic fields higher than about a $\mu$G.
\\
2. The adiabatic invariant is strictly conserved during
intergalactic propagation. Presently it is
not possible to say whether this assumption is true 
or not, due to our very poor knowledge of intergalactic
fields.
\subsection{ Galactic propagation in the extragalactic scenario}
If the concentration factor $b_{\mathrm{conc,equ}}$
is very large, the resulting cosmic-ray density can become
too large to be confined in the Galactic magnetic field.
Cosmic rays then mainly escape through 
additionally formed Parker loops
which are unconnected to the intergalactic magnetic
field. As a limiting case one expects an equilibrium situation 
in which standard Galactic cosmic-ray propagation
remains approximately valid. {\it In this case the only difference to
the standard picture is that hadronic
cosmic rays are supplied from
the outside instead from Galactic sources.}
\\
One can then show that the effective concentration
factor $e$ is given as:
\\
(2) $e = {{d c} \over { D_{\mathrm{G}}}} \simeq 3 \cdot 10^4 $ 
\\
here $ D_{\mathrm{G}}$ is the diffusion coefficient
of Galactic propagation, $d$ is the linear size of the
confinement volume and $c$ is the speed of light.
\subsection{The origin of extragalactic cosmic rays}
It is a completely open question presently
if the intergalactic hadronic cosmic-ray density
was mainly produced in normal galaxies or
active objects. If the latter dominate, radio galaxies
like Cen A seem to be the most natural accelerators
for the locally observed hadronic cosmic rays
in the present scenario. In the local supergalaxy 
it seems quite possible that the intergalactic
medium at the lobes is chemically processed.
{\it But also a supernova-remnant origin 
in normal or ``bright phase'' galaxies is an attractive possibility!}
\section{The counterarguments against an extragalactic
origin of hadronic cosmic rays}
The idea of an extragalactic origin has long been unpopular
because of strong arguments against it.
I will now briefly discuss why all of these arguments loose
their strength if ``flux trapping'' is assumed to operate.
\subsection{\it The expected energy density of 
intergalactic cosmic rays is too small}
A generally
accepted estimate for the density of intergalactic 
hadronic cosmic rays $\rho_{eg}$
expected from normal galaxies
and active objects is about 10$^{-4}$ of the local value $\rho_{loc}$
(Ginzburg, 1993).
Remarkably one gets as a natural and
``untuned'' consequence of our scenario:
$\rho_{\mathrm{loc}}$ $\simeq$ $\rho_{\mathrm{eg}}$ $\cdot$ $e$
(see eq.(2)).
The predicted enhancement factor $e$
has the right order of magnitude to explain the
local energy density of cosmic rays, {\it under the assumption 
of an extragalactical cosmic-ray energy density which is
considered likely by most workers in the field}.
\begin{figure}[hbt]
\centering
\epsfig{file=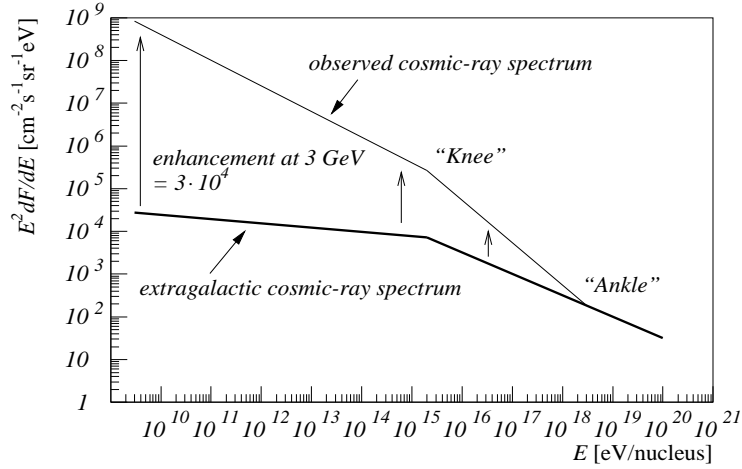,width=11cm,height=8cm,clip=,angle=0}
         \caption{\label{originfigure3.eps}
          An extragalactical scenario for the origin 
          of the observed cosmic-ray spectrum near earth. Plotted
           is the differential flux of hadronic cosmic rays
           multiplied by E$^2$ as a function of the cosmic-ray energy
           per nucleus E . The thin line is a schematic representation
          of experimental results. They are explained by an
          extragalactical spectrum outside the Galaxy (thick line) which
           is enhanced in the Galaxy by
          a factor $e$ (symbolized by the arrows). This factor
           varies $\sim$ E$^{-0.5}$. Only at energies above the
           ``ankle'' the extragalactical spectrum is equal to the
           observed spectrum.}
\end{figure}
More detailed
figure \ref{originfigure3.eps} shows a speculative explanation of the
observed cosmic-ray spectrum under the assumption
of an extragalactic origin.
Here it was taken into account that the enhancement factor $e$
is expected to be energy dependent in the standard theory
of cosmic-ray propagation. While the ``knee'' remains enigmatic
in all scenarios, the position of the ``ankle'' follows as a natural
consequence of the extragalactic scenario.
\subsection{\it There is no $\gamma$ radiation from the SMC}
Sreekumar et al. (1993) 
have argued that their
non-detection of $\gamma$-radiation above 100 MeV
from the Small Magellanic cloud (SMC) rules out an extragalactic
origin of cosmic-rays.  
Their argument is based on the assumption
that the cosmic-ray density in the SMC is equal to the
local Galactic one in extragalactic scenarios.
This assumption does no longer hold in the present special
extragalactic scenario: because of the small
size and dynamically disintegrating state of the SMC,
the ``concentration factor'' for this galaxy is expected to be smaller
than the one for our Galaxy. In fact this 
expected smaller concentration
factor is effectively also the main explanation for the low 
$\gamma$-ray luminosity in the standard ``local''
explanation for cosmic-ray origin. 
\subsection{\it The chemical and isotopical composition
of the hadronic cosmic radiation is very similar to
the one of the solar system, this is in favor
of a local Galactic origin}
The measured abundance pattern is thought to be
determined by general nucleosynthetic principles
which probably operate in all galaxies.
It is possible (though not necessary)
that chemically processed plasma is accelerated in
extragalactic sources.
Moreover in a Galactic scenario 
cosmic rays are ``young'' (about 10$^7$ years) whereas
the solar system is about 5 billion years old. Because of this, the
very good consistency of cosmic-ray and solar-system
abundances is {\it not} a priori expected in a Galactic scenario
(Fields et al.,1993).
\subsection{\it Cosmic-ray clocks measure a
small age of cosmic rays, thus the heavy ions
cannot be of extragalactic origin}
Cosmic-ray clocks like $^{10}$Be
indicate an ``age'' of cosmic rays of a few tens of million
years. 
This is much less
than the expected time since acceleration 
in an extragalactic scenario, which is on the order of
the age of the universe t$_{\mathrm{U}}$ $\simeq$ 1.5 $\cdot$
10$^{10}$ years. This might be interpreted as evidence
against an extragalactic origin.
What cosmic-ray clocks
measure, however, is the time since they propagate in a medium
dense enough to lead to
nonnegligible spallation processing (the radioactive $^{10}$Be
is a spallation product 
from nuclear reactions during propagation
and not a remnant from the acceleration site).
If extragalactic cosmic rays were accelerated in regions with
low matter density (like giant radio lobes) or left the
acceleration site on a very small time scale
(like in an early galaxy with a strong galactic wind)
and then propagated in the intergalactic
medium which has a very low ambient density, the measured
``age'' merely measures the time since entering
the Galaxy, which is on the order of 
the confinement time t$_{\mathrm{conf}}$, like
in the Galactic scenario of cosmic-ray origin.
\subsection{It's a priori unlikely that the electrons
and hadrons in the cosmic radiation have a completely
different origin}
There could be a plausible 
physical mechanism for this ``breaking of the
tie'': as opposed to hadrons,
electrons cannot propagate over cosmological distances
due to inverse Compton scattering on the 3 K$^o$ background
at the energies of interest here.
It is also remarkable that observations of electrons
in several supernova remnants seem to indicate a ``break off'' in
energy spectrum around 10 TeV (Allen, 1997),
in good agreement with an expected maximum
energy of about 100 TeV for older SNRs (both for electrons
and hadrons). There is no evidence
for any sharp feature in hadronic cosmic rays up to much
higher energies (the ``knee'' at 2 PeV).
\vspace{-0.1cm}
\section{Conclusion}
The assumptions made for the basic hypothesis 
of ``flux trapping'' are speculative and 
perhaps controversial. 
A better understanding of very complex magnetohydrodynamic
processes in intergalactic space
is needed  to make
a firm decision whether they are realistic or not.
From a purely phenomenological point of view the
following point of view seems interesting:
\\
{\it Galactic SNRs produce the observed 
electron cosmic-ray flux. The required 
acceleration efficiency is modest
and the high-energy cutoff lies in the region
of 10 TeV, in good agreement with theoretical
expectation 
(see e.g. Mastichiades and de Jager (1996)).
Protons and nuclei are perhaps accelerated with a 
roughly similar
efficiency (i.e. much less than with a 100 times higher
efficiency ) and high-energy
cutoff, and therefore produce a
local hadronic cosmic-ray intensity comparable
to the electron intensity (on the order of 1 $\%$ of the total
intensity below the cutoff). The main part of hadronic cosmic rays
is due to intergalactic cosmic radiation 
which has
an enhanced density in the Galactic confinement
volume.}
\\
I thank I.Holl, D.Petry, S.Pezzoni and H.V\"olk for helpful
discussions and preparation of figures.
\vspace{-1.0cm}
\begin{refs}
\item Allen,G.E.,et al., Proc. 25th ICRC (Durban),4,449 (1997).
\item Baade,W., Zwicky,F., Proc. Nat. Acad. Sci., 20,259 (1934).
\item Berezhko,E.G., V\"olk,H.J.,1997, preprint MPI-H-V12-1997,
to appear in Astropart. Phys. .
\item Berezinskii, V.S. et al.,1990, `Astrophysics
of Cosmic Rays', North Holland, Amsterdam.
\item Brecher,K., Burbidge,G.R., Astrophys.J. 174,253 (1972).
\item Burbidge,G., Phil. Trans. R. Soc. A 277,481 (1974). 
\item Erlykin,A.D.,Wolfendale,A.W.,Zhang,L. and Zielinska,M., A+A 
Suppl.Ser. 120,397 (1996).
\item  Erlykin,A.D.,Smialkowski,A., and Wolfendale,A.W.,
 Proc. 25th ICRC (Durban),3,113 (1997).
\item Fields,B.D., Schramm,D.N.,Truran,J.W., 
Proc. 23th ICRC (Calgary),2,398 (1993).
\item Ginzburg, V.L., Phys. Usp. 36,587 (1993).
\item He\ss \,M., Proc. 25th ICRC (Durban),3,229 (1997).
\item Lessard,R.W. et al., Proc. 25th ICRC (Durban),3,233 (1997).
\item Mastichiadis,A., de Jager,O.C.,1996, submitted to Astron. Astrophys.,
astro-ph 9606014.
\item Plaga, R., ``An extragalactic ``flux trapping'' origin of the dominant
part of hadronic cosmic rays?'', accepted for publication
in Astron. Astrophys., astro-ph/9711094.
\item Sciama,D., Mon.Not.R.Ast.Soc. 123,217 (1962).
\item Sreekumar,P., et al., Phys.Rev.Lett. 70,127(1993).
\item Tanimori,T. et al., IAU telegram 6706 (1997).
\end{refs}
\pagebreak

\end{document}